\documentstyle[aaspp4,12pt]{article}
\begin{document}

\def\P{\bar{\Phi}}
\def\r{r}
\def\alpha{C}
\def\mu{\lambda}
\def\Sigma{\sigma}
\def\sles{\lower2pt\hbox{$\buildrel {\scriptstyle <}
   \over {\scriptstyle\sim}$}}
\def\sgreat{\lower2pt\hbox{$\buildrel {\scriptstyle >}
   \over {\scriptstyle\sim}$}}
\newcommand{\ul}{\underline{\hspace{40pt}}}

\title{Self-Similar Collapse of Nonrotating Magnetic
Molecular Cloud Cores}

\author{Ioannis Contopoulos, \altaffilmark{1}
Glenn E. Ciolek, and Arieh K\"{o}nigl}
\affil{Department of Astronomy and Astrophysics,
University of Chicago \\
5640 S. Ellis Avenue, Chicago, IL 60637
\\
\vspace{3ex} Accepted for publication in {\it The Astrophysical Journal}, 1 Sept 1998 issue
}
\altaffiltext{1}{Present address: Physics Department, University of
Crete, P. O. Box 2208, Heraklion 71003, Greece}

\begin{abstract}
We obtain self-similar solutions that describe the
gravitational collapse of nonrotating, isothermal, magnetic
molecular cloud cores.  We use simplifying assumptions but explicitly
include the induction equation, and the semianalytic
solutions we derive are the first to account for the effects of ambipolar
diffusion following the formation of a central point mass. Our
results demonstrate that, after the protostar first forms,
ambipolar diffusion causes the magnetic flux to decouple in a growing region
around the center. The decoupled field lines remain approximately stationary
and drive a hydromagnetic C-shock that moves outward at a
fraction of the speed of sound (typically a few tenths of a kilometer per
second), reaching a distance of a few thousand AU at the end of
the main accretion phase for a solar-mass star.  We also show
that, in the absence of field diffusivity, a contracting core
will not give rise to a shock if, as is likely to be
the case, the inflow speed near the origin is nonzero at the
time of point-mass formation. Although the evolution of
realistic molecular cloud cores will not be exactly self
similar, our results reproduce the main qualitative features
found in detailed core-collapse simulations (Ciolek \& K\"{o}nigl 1998).

\end{abstract}

\newpage
\section{Introduction}

Low-mass stars are generally believed to form as a result of the
gravitational collapse of molecular cloud cores. The cores are
initially supported by thermal and magnetic forces, but because
of ambipolar diffusion (the drift of ions, to which the magnetic
field lines are attached, relative to the dominant neutral gas
component), they gradually lose their magnetic support and
eventually collapse after becoming ``supercritical''
(see, e.g., Mouschovias 1987 for a review).\footnote{In
this paper we reserve the term ``core'' for the high-density central region of
a molecular cloud and do {\em not} apply it to the point mass
that forms from the collapse of such a core.} The most detailed
numerical treatments to date of the problem of the ambipolar
diffusion-initiated formation of supercritical cores and the early
stages (prior to point mass formation) of their subsequent
dynamical collapse have been presented by Mouschovias and collaborators
(Fiedler \& Mouschovias~1992, 1993; Ciolek \& Mouschovias~1993, 1994, 1995,
hereafter CM93, CM94, CM95; Basu \& Mouschovias~1994, 1995a, 1995b, hereafter
BM94, BM95a,b). Because the timescale for core formation is much
longer than the timescale for dynamical collapse,
special numerical techniques had to be employed in
these calculations. The simulations were terminated when the
central densities reached $\sim 10^{10}\ {\rm cm}^{-3}$ and the
underlying assumptions of isothermality (e.g., Gaustad~1963)
and flux freezing onto the ions (e.g., Pneuman \& Mitchell~1965)
broke down.  These
calculations were nevertheless able to demonstrate that 
{\em supercritical cores begin to collapse dynamically
before a point mass (i.e., a protostar) appears at the origin}.

The dynamical evolution of supercritical cores after their formation
has been studied by many researchers. Solutions exist for the collapse of
nonrotating, self-gravitating spheres without thermal support (Henriksen~1994),
self-gravitating spheres with thermal support (Penston~1969;
Larson~1969; Shu~1977; Hunter~1977; Boss \& Black~1982; Whitworth \&
Summers~1985; Foster \& Chevalier~1993) as well as with a
combined thermal and isotropic magnetic pressure support (Chiueh \& Chou 1994),
and self-gravitating disks with thermal support (Narita, Hayashi, \&
Miyama~1984; Matsumoto, Hanawa, \& Nakamura~1997) and also with
ordered, frozen-in magnetic fields (Nakamura, Hanawa \&
Nakano~1995; Li \& Shu~1997, hereafter LS). In order to
choose a particular solution for a given problem, one needs to know
the properties of the supercritical core at the time of its
formation. This information, however, can only be gleaned
from a study of the preceding, quasi-static evolution of the
core under the influence of ambipolar diffusion.
Although different assumptions about the initial state of the
core yield solutions that are qualitatively similar in their gross behavior
(the core collapses with near free-fall speeds and a point mass
eventualy forms at the center), the solutions do differ in
such important details as the accretion rate onto the
central point mass and the formation (or absence) of shocks.

The well-known examples of the Larson-Penston (1969) and
Shu~(1977) similarity solutions in fact represent two extremes
of a whole continuum of self-similar collapse solutions specified by a
cloud's initial configuration and the conditions at its boundary
(Hunter 1977; Whitworth \& Summers 1985; see also Chiueh \& Chou 1994
for a generalization to the case of an isotropic internal magnetic pressure).
The Larson-Penston (1969) solution is characterized by a
spatially uniform, supersonic
(at $\sim 3.3$ times the isothermal speed of sound $C$) infall
speed and an inverse-square dependence of the density $\rho$ on
the radius $r$ at the instant of point-mass formation (PMF); the
mass accretion rate at the center is $\sim 29 \
C^3/G$ (where $G$ is the gravitational constant) at that instant and
increases to $\sim 47 \ C^3/G$ immediately after PMF. Numerical
simulations of the collapse of nonmagnetic isothermal spheres (Hunter
1977; Foster \& Chevalier 1993) have indicated that this
solution provides a good approximation to the conditions near
the center at the PMF epoch for clouds that are initially near a
marginally stable equilibrium. The Shu (1977) solution strictly applies
only to the post-PMF evolutionary phase: it consists of an
inner free-fall region and a hydrostatic outer envelope that are
separated by an outward-propagating (at a speed $C$ relative to
the gas) expansion wave.
The envelope corresponds to a singular isothermal sphere
($\rho \propto r^{-2}$) and the mass accretion rate onto the
center is $\sim 1 \ C^3/G$. In applying this solution to real
systems, it was proposed to identify the initial core
configuration at the end of the quasi-static ambipolar-diffusion
phase with a singular isothermal sphere (or, more generally, a
toroid) at the instant of PMF (e.g., Shu, Adams, \& Lizano 1987;
Li \& Shu
1996). However, as we noted above, the conclusion from detailed numerical
simulations has been that the dynamical phase of core collapse
generally commences well before the PMF epoch, so that
the innermost region is not well represented by a
quasi-static solution at the time of point-mass formation.

Another interesting effect that depends on the specific
choice of initial conditions and on the detailed physical
properties of the collapsing core is the formation (or absence)
of shocks (e.g., Tsai \& Hsu 1995; LS).  For example, LS discovered
that when, instead of a spherical core, one considers the collapse of
a flattened disk, the expansion wave of Shu~(1977) becomes a shock.
As we show in this paper, when one takes proper account of the fact that
supercritical cores collapse dynamically before a point mass
first forms at the origin, that shock disappears.  Nevertheless,
a physical basis for the formation of shocks in collapsing
magnetized molecular cloud cores has been discussed by Li \&
McKee (1996), who argued that a hydromagnetic C-shock will
appear as a result of the outward diffusion of inwardly advected
magnetic flux. The existence of such a shock has been
confirmed in the numerical simulations of Ciolek \& K\"{o}nigl
(1998, hereafter CK), and it is, in fact, a salient feature of the
semianalytic solutions derived in this paper.

The aim of the present work is to clarify the effects of
ambipolar diffusion in dynamically collapsing supercritical
cores. Toward this goal, we construct semianalytic,
time-dependent similarity solutions of gravitationally
contracting, magnetized, isothermal disks. Although the
evolution of real molecular cloud cores is not expected to be
exactly self similar, we demonstrate, through a comparison
with the detailed numerical simulations of CK, that our
solutions capture the main traits exhibited by the latter
calculations. Based on an analogous comparison with the results
of numerical simulations, Basu (1997) showed that a self-similar
scaling describes the pre-PMF evolution in the innermost
flux tubes of collapsing supercritical cores quite well.
To complement his study, we concentrate in this paper on the
post-PMF evolutionary phase.
Our approach differs, however, from that of Basu
(1997) in that we explicitly solve the induction equation,
whereas he accounted for the effects of ambipolar diffusion only
in a phenomenological manner.\footnote{Our work is thus also
distinguished from that of Safier, McKee, \& Stahler (1997), who
studied the effects of ambipolar diffusion in the spherically
symmetric, quasi-static limit without explicitly solving the
induction equation.} In fact, the solutions that we derive,
while involving various simplifications, are nevertheless
the first to consistently incorporate ambipolar diffusion into a
self-similar representation
of the collapse of a magnetized cloud core.
\footnote{The effect of {\em weak} magnetic fields
on a dynamically collapsing core in the presence of ambipolar diffusion
was previously investigated
by Galli \& Shu (1993a), who carried out a perturbation
expansion of the (nonmagnetic) spherical similarity solution of
Shu (1977). As was already noted and discussed by Li \& McKee
(1996), the semianalytic solution derived in that paper, as well
as the associated numerical
calculation in Galli \& Shu (1993b), did not uncover the
existence of a flux diffusion-driven shock.} We formulate the
problem in \S 2, present our solutions in \S 3, and discuss the
results in \S 4. Our conclusions are summarized in \S 5.

\section{Mathematical Formulation and Approximations}
  
The problem of the self-initiated formation and contraction of cloud
cores due to ambipolar diffusion in axisymmetric, self-gravitating,
isothermal, magnetic molecular clouds was formulated in detail in
CM93. In the presence of an ordered, large-scale magnetic field the
contracting cloud core assumes a disk-like configuration on account of
the fact that the magnetic stresses inhibit motion normal to the field
lines (see Fiedler \& Mouschovias 1993). Along the field lines it is a
good approximation to assume that the cloud is at all times in
hydrostatic equilibrium, with thermal pressure providing support
against vertical gravity. For simplicity, we neglect both the pressure
support provided by internal hydromagnetic waves (i.e.,
``turbulence'') and the vertical squeezing of the core by the external
thermal and magnetic pressure
(including, in particular, the
squeezing induced by the radial field component at the disk
surface; see eq. [9] in CK).
For a self-gravitating disk with
possibly a point mass at the center, the hydrostatic balance equation
can thus be written as
\begin{equation}
\rho C^2 = \frac{\pi}{2}G\sigma^2
+\frac{GM_c\rho h^2}{2r^3}\ ,
\label{ver}
\end{equation}
where $r$ is the cylindrical radius, $\sigma$ is the surface density,
$\rho$ is the gas density,
$h\equiv \sigma/2\rho$ is the vertical scale height (assumed to
be $\ll r$), $C=(k_B T/m_n)^{1/2}$ is the isothermal speed of
sound (with $k_B$ being the Boltzmann constant, $T$ the temperature, and
$m_n$ the mean mass of a gas particle), and $M_c$ is the mass of the
star that forms at the center.
Equation~(\ref{ver}) implies that
\begin{equation}
h= \left\{ \begin{array}{ll}
 rC(GM_c/2r)^{-1/2} & 
\mbox{near the central point mass} \\
C^2[\pi G\Sigma]^{-1} & \mbox{far from the central point mass}\ .
\end{array}
\right.
\label{h}
\end{equation}
The collapse of the core can be described using the mass and
(radial) momentum
conservation equations, which, in cylindrical coordinates $(\r,\phi,z)$, read
\begin{equation}
\frac{\partial\sigma}{\partial t}+
\frac{1}{\r}\frac{\partial}{\partial\r}(\r\sigma u) = 0
\label{sigma}
\end{equation}
and
\begin{equation}
\frac{\partial u}{\partial t}+
u\frac{\partial u}{\partial \r} = g_\r
-C^2\frac{\partial \ln\sigma}{\partial \r}
+\frac{B_z}{2\pi\sigma}\left(B_{r,s}
-h\frac{\partial B_z}{\partial r}\right)\ ,
\label{acc}
\end{equation}
respectively. Here $B_{\r,s}$ is the radial component of the magnetic
field at the surface of the thin disk (by symmetry, $B_\r=0$ at the
midplane $z=0$, and $\partial B_\r/\partial z
\approx B_{\r,s}/h$), $u$ is the radial velocity component, and $g_\r$ is
the radial component of the gravitational acceleration.  As is well
known, the calculation of $g_\r$ for a self-gravitating thin
disk is in general nontrivial.  The above equations are
supplemented by the induction equation for the magnetic field
(see derivation below),
\begin{equation}
\frac{\partial \Psi}{\partial t}=
-2\pi \r \left[u B_z +
C \frac{\eta}{\mu^2}
\left(\frac{\pi G\sigma h}{C^2}\right)^{1/2}
\left(B_{\r,s}-h\frac{\partial B_z}{\partial r}\right)
\right]\ ,
\label{induction}
\end{equation}
where
\begin{equation}
\eta \equiv \tau_{ni}/(8\pi G \rho)^{-1/2} \approx 0.7\
\xi_{-17}^{-1/2}\ ,
\label{ionize}
\end{equation}
the ratio of the mean collision time $\tau_{ni}$ of a
neutral particle with a sea of ions and the approximate
free-fall time due to the disk self-gravity, is a measure of the
efficiency of ambipolar diffusion in the disk, and
\begin{equation}
\lambda\equiv \frac{2\pi G^{1/2}\sigma}{B_z}
\end{equation}
is the dimensionless local mass-to-flux ratio. In equation (\ref{induction})
$\Psi(\r,t)\equiv \int_0^\r 2\pi \r' B_z(\r',t) {\rm d}\r'$ is
the magnetic flux threading the disk interior to the radius $\r$
at time $t$, whereas in equation~(\ref{ionize})
$\xi=10^{-17}\xi_{-17}\ {\rm s}^{-1}$ is the cosmic-ray ionization
rate per hydrogen nucleus (e.g., Watson 1976).
In deriving equation (\ref{induction}) we
have assumed that the disk is weakly ionized and that the
magnetic field is ``frozen'' into the ions (valid for disk
densities $\la 10^{11}\ {\rm cm}^{-3}$). The first assumption implies
\begin{equation}
\rho_i\ll \rho_n\ ,
\label{uD}
\end{equation}
where the subscripts $i$ and $n$ refer, respectively, to the
ions and neutrals. Under this assumption, one can identify $\rho$
in the preceding equations with $\rho_n$, and $u$ with $u_n$.
The speeds $u_n$ and $u_i$ are not, in general,
equal, and their difference $u_D\equiv (u_i -  u_n)$ is referred
to as the (radial) ion--neutral drift speed.
The second assumption can be expressed by writing the
flux conservation equation in the form
\begin{equation}
\frac{\partial \Psi}{\partial t}=
-2\pi ru_i B_z=-2\pi r (u_n+u_D)B_z\ .
\label{ionind}
\end{equation}
Magnetic forces that act on the ions are transferred to the neutrals through
a collisional drag force,
\begin{equation}
\frac{\rho u_D}{\tau_{ni}}=\frac{B_z}{4\pi}\left(-\frac{\partial
B_z}{\partial r} +\frac{\partial B_\r}{\partial z}\right)\ .
\label{drift}
\end{equation}
For a disk that
consists of a molecular hydrogen gas with a helium number
density that is 10\% of the density of hydrogen nuclei, the
neutral--ion mean collision time is given by
$\tau_{ni}=1.4(m_i+m_{H_2})/[\rho_i <\sigma w>_{iH_2}]$, where
$<\sigma w>_{iH_2}= 1.69\times
10^{-9}\ {\rm cm}^{3}\ {\rm s}^{-1}$ is the average collisional rate
between ions and hydrogen molecules (e.g., McDaniel \& Mason 1973). 
By balancing ionization by cosmic
rays with dissociative recombination, one can express the ion
number density as $n_i\approx (\xi n/a_{\rm dr})^{1/2}$ (Elmegreen
1979), where $a_{\rm dr}\approx 10^{-6}\ {\rm cm}^{3}\ {\rm s}^{-1}$ is the
electron dissociative recombination rate (e.g., Dalgarno~1987)
and $n$ is the neutral particle number density. Although this
expression
(which was used in deriving the dependence of $\eta$
on $\xi$ in eq. [\ref{ionize}])
is strictly valid only for densities $n\la 10^{6}\ {\rm cm}^{-3}$ (at
densities $\gg 10^6~{\rm cm}^{-3}$, $n_i \approx const$; see
Figs.~2$c$ and 4$c$ in CM94, and Figs. 1 -- 3 in Ciolek \&
Mouschovias~1998), we
nevertheless adopt it in this paper for all disk densities
since it allows us to obtain a self-similar set of
equations that can be directly integrated.
Equation~(\ref{induction}) follows from combining
equations~(\ref{ionind}) and (\ref{drift}) and taking the ion
mass to be $m_i=29 \, m_H$. Equation~(\ref{drift}) brings out
the well-known fact that, in order to support a nonuniform
magnetic field distribution in the disk, the
ion--neutral drift speed has to be nonzero. We return to this
basic point in \S 3.1.

Given the distributions $\Sigma(\r,0)$, $u(\r,0)$, and $B_z(\r,0)$ at
some initial time $t=0$, one can in principle integrate
equations~(\ref{sigma})--(\ref{induction}) provided that
$g_\r(\r,t)$ and $B_{\r,s}(\r,t)$ can be expressed as functions
of $\sigma$ and $B_z$, respectively. In the limit of a thin disk
with negligible mass outside,
one can obtain the radial gravitational acceleration at
radius $\r$ by adding the contributions from rings of disk material at
all radii $\r'$,
\begin{equation}
g_\r(\r,t)=-\frac{G}{\r^2}\int_0^\infty
2\pi \r' \Sigma(\r',t){\cal R}(\r'/\r) {\rm d}\r'\ .
\label{gravity}
\end{equation}
In this equation,
\begin{equation}
{\cal R}(X)=\frac{1}{2\pi}
\int_0^{2\pi} \frac{(1-X\cos\phi){\rm d}\phi}
{(1+X^2-2X\cos\phi)^{3/2}}
={\cal K}(X)+X\frac{{\rm d}{\cal K}}{{\rm d}X}\ ,
\nonumber
\end{equation}
where
\begin{equation}
{\cal K}(X)\equiv \frac{2}{\pi (1+X)}K(4X/[1+X]^2)\ ,
\nonumber
\end{equation}
with $K(x)$ being the complete elliptic integral of the first
kind (see CM93). Figure 1 shows a plot of ${\cal R}(X)$ obtained
using approximate formulas for $K(x)$ (e.g.,
Abramowitz \& Stegun~1965, p.~591). It is seen that
contributions to the gravitational acceleration at a radius $\r$ arise
from disk material both interior {\em and} exterior to $\r$.  There
exist, however, two limits in which the simple expression
\begin{equation}
g_\r=-\frac{GM}{\r^2}\ 
\label{simplegravity}
\end{equation}
holds true, where $M(\r,t)\equiv \int_0^\r 2\pi\r' \Sigma(\r',t){\rm
d}\r'$ is the mass interior to the radius $\r$ at time $t$: (a) when
$\sigma\propto \r^{-1}$, and (b) when the gravitational pull of
a point mass at the center dominates over the disk self-gravity.
As we demonstrate below,
equation~(\ref{simplegravity}) is a very good approximation to the true
value of $g_\r$ during the post-PMF phase of the disk
evolution. For future reference, we note here that the mass
conservation equation (\ref{sigma}) can be rewritten in terms of
$M(\r,t)$ in the form
\begin{equation}
\frac{\partial M}{\partial t}+
u\frac{\partial M}{\partial\r} = 0\ .
\label{mass}
\end{equation}

What about $B_{\r,s}$? Assuming that the medium surrounding the
disk is current-free,
the magnetic field can be expressed as the gradient of a
potential: ${\bf B}=\nabla\Phi$. Now, since $\nabla\cdot{\bf B}=0$,
this scalar potential satisfies Laplace's equation
\begin{equation}
\nabla^2\Phi=0 
\end{equation} 
with Neumann boundary conditions
\[ \frac{\partial \Phi}{\partial z}=
\left\{ \begin{array}{ll}
B_z & \mbox{at the surface of the disk ($z\simeq 0$)} \\
B_{z,\infty} & \mbox{at infinity}\ ,
\end{array}
\right.
\]
where $B_{z,\infty}$ is the uniform and constant large-scale
magnetic field that is frequently observed to thread molecular
clouds on scales that are large in comparison with their
(flattened) inner cores (e.g., Hildebrand, Dragovan, \& Novak 1984;
Novak, Predmore, \& Goldsmith~1990; Kane et al.~1993).
Therefore, in the ideal case of a current-free medium
outside a thin disk, the function $\Phi-zB_{z,\infty}$ satisfies
the same equation and (Neumann) boundary conditions as the
gravitational potential $V$ (where ${\bf g}=-\nabla V$,
$\nabla^2 V=0$, $\partial V/\partial z=2\pi G\Sigma$ at $z=0$,
and $\partial V/\partial z= 0$ at infinity). Hence,
by direct analogy with the expression for $g_\r$ in
equation~(\ref{gravity}), one can write
\begin{eqnarray}
B_{\r,s}(\r,t) & \equiv & 
\left. \frac{\partial \Phi}{\partial r}\right|_{z=0}\nonumber \\
 & = & \left. \frac{\partial}{\partial r}
(\Phi-zB_{z,\infty})\right|_{z=0}\nonumber \\
& = & \frac{1}{\r^2}\int_0^\infty \r'
(B_z(\r',t)-B_{z,\infty}){\cal R}(\r'/\r) {\rm d}\r' 
\label{Bra}\\
& \approx &
\frac{1}{\r^2}\int_0^\infty \r' B_z(\r',t){\cal R}(\r'/\r)
{\rm d}\r'\ ,
\label{Brb}
\end{eqnarray}
where the last approximation is valid when\/ $B_z$ at the disk
surface is $\gg B_{z,\infty}$ (see CM93, CM94, CM95, BM94, and
BM95a,b). When one considers {\em only} the inner parts of the
collapsing disk, as in the present analysis, one can
neglect $B_{z,\infty}$.\footnote{When the strong inequality
$B_z\gg B_{z,\infty}$ no longer
holds, as is the case in the outer parts of the supercritical core,
the expressions in equations~(\ref{gravity}) and (\ref{Bra}) differ, and
magnetic tension can indeed ``overwhelm'' self-gravity. This
should clarify the issue raised in the footnote on p. 248 of
LS.}  Pursuing the analogy with
gravity even further, one sees that it is possible to use the simple expression
\begin{equation}
B_{\r,s}=\frac{\Psi}{2\pi\r^2}
\label{simpleBr}
\end{equation}
when either (a) $B_z\propto \r^{-1}$ or (b) the magnetic field advected
to the center can be represented by a split monopole that
dominates over the disk
magnetic field.  We will demonstrate that either (a) or (b) are applicable
in a collapsing disk following the formation of a central point
mass, so that equation~(\ref{simpleBr}) provides a good
approximation to $B_{\r,s}$ during that phase.

\section{Self-Similar Solutions}

\subsection{The ``Pivotal'' State}

Although one can in principle integrate equations
(\ref{sigma})--(\ref{induction}) for any set of physically
reasonable initial conditions [expressed by the functions
$\sigma(\r,0)$, $u(\r,0)$, and $B_z(\r,0)$], one
particular set of initial distributions makes it possible to obtain
self-similar solutions that compare very favorably with
detailed numerical simulations.\footnote{Note that, although the
numerical models exhibit a self-similar behavior at late times
irrespective of the detailed early-time ($t \rightarrow - \infty$)
conditions, only one particular
set of distributions corresponds to exact self-similarity at $t=0$.}
Specifically, the only initial surface-density
and magnetic-field distributions that are consistent
with a self-similar evolution are
\begin{equation}
\sigma(\r,0)\propto B_z(\r,0)=B_{\r,s}(\r,0)\propto \frac{1}{\r}\ .
\label{SB}
\end{equation} 
For these distributions the initial mass-to-flux ratio is constant
throughout the disk:\footnote{LS refer to disks in which the
mass-to-flux ratio is a spatial constant as being {\em isopedic}.}
\begin{equation}
\frac{\sigma(\r,0)}{B_z(\r,0)}\equiv
\frac{\lambda_o}{2\pi G^{1/2}}={\rm const.}
\label{MF}
\end{equation}
As it turns out, the scalings given by equation (\ref{SB}) are
indeed representative of the central region of
a collapsing core at the time of point-mass formation (see CM94, CM95, BM94,
BM95a,b, and Basu 1997). We therefore associate that instant with the time
$t=0$, and, adopting the nomenclature introduced by Li \&
Shu~(1996), we denote the corresponding disk configuration as the
``pivotal'' state of the model.

We emphasize once again the point first made in \S 1 that the
pivotal state is {\em not} the end
state of the sequence of subcritical quasi-static equilibria that
slowly evolve due to ambipolar diffusion.
\footnote{It is interesting to note, however, that our $t=0$
surface-density and magnetic field distributions
(eq. [\ref{SB}]) are the same as those of the static pivotal
state in the LS similarity solution, which corresponds to a singular
isothermal disk (see Li \& Shu 1996).}
In fact, our pivotal
state represents a dynamically collapsing disk with
\begin{equation}
\dot{M}(r,0)\neq 0\ , \ u(r,0)\neq 0\ .
\end{equation}
A simple physical argument clarifies why the self-similar
pivotal state cannot correspond to hydrostatic equilibrium when
ambipolar diffusion
is self-consistently taken into account.  Self-similarity requires a
nonuniform distribution of $B_z$ at $t=0$ (eq.~[\ref{SB}]), which, in
order to be supported, requires a {\em nonzero (and positive) drift
speed $u_D$} between ions and neutrals (eq.~[\ref{drift}]);
otherwise $B_{\r,s}(\r,0)$ would have been zero and
$B_z(\r,0)$ would
have been uniform, which is clearly not the case.  During the initial
phase of core collapse ($t \leq 0$) the magnetic flux either
remains fixed in space or else is advected
inward, i.e.  $u_i(\r,t\leq 0) \la 0$. Hence
\begin{equation}
|u(\r,0)|\equiv
|u_i(\r,0)-u_D| \ga u_D = C \frac{\eta}{\lambda_o^2}
\left(1+\frac{h}{r}\right)\ ,
\label{u}
\end{equation}
where the expression for $u_D$ follows from equations (\ref{h}),
(\ref{drift}), (\ref{SB}), and (\ref{MF}). One can obtain an
order-of-magnitude estimate for $\dot{M}(r,0)$ by using also
equation (\ref{acc}) to approximate $\sigma(r,0)$ by its 
value in equilibrium,
\begin{equation}
\sigma(\r,0)\approx \frac{C^2}{2\pi G\r}\left(
\frac{\lambda_o^2+2}{\lambda_o^2-1}\right)\ ,\ 
u_D\approx \frac{3\eta C}{(\lambda^2_o+2)}\ , \
{\rm and}\ 
\dot{M}(\r,0)\approx
\frac{C^3}{G}\frac{3\eta} {(\lambda_o^2-1)}\ .
\label{S}
\end{equation}
Taking as representative values $\lambda_o=2.9$ and
$\xi_{-17}=5$ (see \S 3.4), one infers
\begin{equation}
|u(\r,0)|\ga 0.1 C\ {\rm and}\ \dot{M}(\r,0)\approx 0.2
\left(\frac{T}{10\ {\rm K}}\right)^{3/2}\ {\rm M}_{\odot}\
{\rm per}\ 10^6\ 
{\rm yr}\ ,
\label{numbers}
\end{equation}
where $T$ is the temperature.  The numerical simulations of CK suggest
that, in fact, $|u(\r,0)|\gg u_D$, and that $\dot{M}(\r,0)$ is of the
order of a few solar masses per $10^6$ years.\footnote{Note that, for a
$10\ {\rm K}$ $H_2$ gas with an interstellar
He abundance, $m_n = 2.33~\rm{a.m.u.}$ and $C^3/G=1.58\
{\rm M}_{\odot}\ {\rm per}\ 10^6\ {\rm yr}$.}

\subsection{Dimensionless Equations}

We start by introducing a similarity variable $x$ and dependent nondimensional
variables $\hat{h}(x)$, $a(x)$, $v(x)$, $g(x)$,
$m(x)$, $\dot{m}(x)$, ${\bf b}(x)$, and $\psi(x)$:
\begin{equation}
x=\r/C t\ ,
\end{equation}
\begin{equation}
h= C t\, \hat{h}(x)\ , \ \Sigma(\r,t)=[C/(2\pi Gt)]\, a(x)\ , \
u(r,t)=C \, v(x)\ ,\ g_\r=(C/t)\, g(x)\ ,\
\label{nondimensional}
\end{equation}
\begin{equation}
M(\r,t)=(C^3 t/G)\, m(x)\ , \ \dot{M}(\r,t)=(C^3 /G)\,
\dot{m}(x)\ ,
\end{equation}
\begin{equation}
{\bf B}=(C/[G^{1/2}t])\, {\bf b}(x)\ , \ \Psi(\r,t)=(2\pi C^3
t/G^{1/2})\, \psi(x)\ , 
\end{equation}
where
\begin{equation}
m(x)=\int_0^x a(x')x' {\rm d}x'\ , \
\dot{m}(x)=-axv\ , \ 
\psi(x)=\int_0^x b_z(x')x' {\rm d}x'\ .
\label{mpd}
\end{equation}

In these variables, equations~(\ref{sigma})--(\ref{induction}) and
(\ref{ver}) can be rewritten as
\begin{eqnarray}
\frac{{\rm d} a}{{\rm d}x} & = &
\frac{a}{1-(x-v)^2}\left[
g+ \frac{b_z}{a}\left(b_{\r,s}-\hat{h}\frac{{\rm d}b_z}
{{\rm d}x}\right)+\frac{(x-v)^2}{x}\right]
\label{sys1}\\ 
\frac{{\rm d}v}{{\rm d}x} & = &
(x-v)\frac{{\rm d}\ln a}{{\rm d}x}+\frac{x-v}{x}\ ,
\label{sys2}\\ 
\psi & = & 
x(x-v)b_z-x\eta \left(\frac{b_z}{a}\right)^2
\left(\frac{a\hat{h}}{2}\right)^{1/2}
\left(b_{\r,s}-\hat{h}\frac{{\rm d}b_z}{{\rm d}x}\right)\ ,
\label{sys3}\\
\hat{h} & = & 
\frac{ax^3}{2m_c} \left [-1+ \left (1+\frac{8m_c}{a^2 x^3}
\right)^{1/2} \right ]\ ,
\label{sys4}
\end{eqnarray}
where
\begin{equation}
g=-\frac{1}{x^2}\int_0^\infty
 x' a(x'){\cal K}(x'/x) {\rm d}x'\ ,\ 
b_{\r,s}=\frac{1}{x^2}\int_0^\infty
 x' b_z(x'){\cal K}(x'/x) {\rm d}x'\ ,\ 
\label{correct}
\end{equation}
and $m_c\equiv m(0)$. These equations are integrated subject to the
initial conditions
\begin{equation}
a \rightarrow \frac{A}{x}\ ,\ b_z \rightarrow \frac{a}{\lambda_o}\ ,\ v
\rightarrow v_o\ ,\ \dot{m}\rightarrow Av_o\equiv \dot{m}_o\ \ {\rm
as}\ x\rightarrow +\infty\ .
\label{initial}
\end{equation}
Since the value ($m_c$) of the central point mass is not known a
priori, the solution requires a numerical iteration to calculate
$\hat{h}$.

The complicated expressions for $g$ and $b_{\r,s}$ in
equation~(\ref{correct}) can in principle be solved by means of an
iterative procedure. As noted by LS, it is natural to start such an
iteration using the monopole terms
\begin{equation}
g\approx -\frac{m}{x^2}\ , \ b_{\r,s}\approx \frac{\psi}{x^2}\ ,
\label{mpsi}
\end{equation}
where, by virtue of equation (\ref{mass}), $m(x)$ satisfies
\begin{equation}
m=x(x-v)a\ .
\label{redmass}
\end{equation}
With these substitutions one can solve for $a(x)$ and $b_z(x)$, and,
in turn, use the latter distributions to improve the estimates for
$g(x)$ and $b_{\r,s}(x)$. This procedure can be repeated until a good
convergence is attained. As it turns out, the behavior of the disk at
$t\geq 0$ is already very well described by the monopole
approximation, so, in what follows, we only use the monopole terms in
deriving our solutions.  

The system of equations
(\ref{sys1})--(\ref{sys4})
contains a singular line at the locus of
points in the ($x,-v$) plane where 
\begin{equation}
(x-v)^2=1\ ,
\label{singular}
\end{equation}
which corresponds to the sonic line. This result should be
contrasted with the criticality
condition for ideal MHD, which involves the magnetosonic speed
instead of just the thermal sound speed (see \S 3.3).
The mathematical reason for this difference is that magnetic
diffusivity introduces derivatives of higher
order than in ideal MHD (e.g., Ferreira \& Pelletier 1995). There is
also an interesting physical explanation of this result. In a
magnetohydrodynamic flow, information propagates via magnetosonic and
sonic waves.  The critical surfaces that appear in steady-state problems
can be thought of as being the relics of the
time-dependent problem in which the various waves had sufficient time
to propagate to the exterior boundaries and communicate the
information associated with those boundaries to the whole flow
(Blandford \& Payne~1982).  When diffusivity is
present, the only waves that can survive propagation to and from
infinity are the sonic waves (since the magnetosonic waves simply dissipate
away): this is the physical reason why only the sonic critical point
appears in steady-state, diffusive MHD. The collapse
problem that we have formulated,
although time dependent, resembles a steady-state problem in
this respect because of the restrictive assumption of
self-similarity, which effectively combines time and space into
a single variable.  It therefore also involves the establishment of a
critical surface, whose nature is determined by the above argument.

\subsection{The Flux-Frozen Case}

We first consider the ideal-MHD case $\eta=0$, with $v_o\neq
0$. In this limit, one can rewrite equation~(\ref{sys3})
(using eqs. [\ref{mpd}] and [\ref{sys2}]) as
\begin{equation}
\frac{{\rm d}\ln b_z}{{\rm d} x} =
\frac{{\rm d}\ln a}{{\rm d} x}\ ,
\end{equation}
which introduces a singular line in the ($x,-v$) plane given by
\begin{equation}
(x-v)^2=1+2\lambda^{-2}
\label{ideal}
\end{equation}
(where, under the assumed ideal-MHD conditions, the mass-to-flux ratio
$\lambda=a/b_z$ is a constant). This singular line is {\em different} from
the one given by equation (\ref{singular}) for the nonideal case. In
physical units, equation (\ref{ideal}) can be rewritten as $r/t
-u=\pm(C^2+u_{\rm A}^2)^{1/2}$, where $u_{\rm A}\equiv B_z/(4\pi[\pi
G\Sigma^2/2C^2])^{1/2}$ and $(C^2+v_{\rm A}^2)^{1/2}$ are,
respectively, the effective {\em Alfv\'{e}n} and {\em magnetosonic}
speeds in the disk.\footnote{When the density is given by $\rho=\pi
G\Sigma^2/2C^2$ (see eq.~[\ref{ver}]), $u_{\rm A}$ is equal to the usual
Alfv\'{e}n speed in the disk. The expression $(C^2+u_{\rm
A}^2/2)^{1/2}$ given by Shu \& Li (1997, footnote 2) for the effective
magnetosonic speed is evidently an error.} The equations now describe
the $t>0$ collapse of an isothermal and isopedic disk with
$\lambda(x)=\lambda_o=$const. The solution for the representative parameter
set $\lambda_o=2.9$, $-v_o=1$, and $\dot{m}_o=3$ (see \S 3.4) is shown in
Figures~$2a$--$2d$. It is seen that, as the collapse progresses,
mass accumulates at the origin at a rate
\begin{equation}
\dot{M}_c= 9.6\ 
\left(\frac{T}{10\ {\rm K}}\right)^{3/2}\ {\rm M}_{\odot}\ 
{\rm per}\ 10^6\ {\rm yr}\ ,
\end{equation}
corresponding to $m_c=6.1$ in the expression
\begin{equation}
M_c(\r=0^+,t) =  (C^3 t/G) \ m_c
\label{pointmass}
\end{equation}
for the central point mass.
Note that, as discussed in connection with previous collapse calculations
(see \S~1), there is a significant and rapid increase ($\dot{m}_c = 2.0~\dot{m}_o$) in
the accretion rate following point-mass formation.
Near the origin, the flow proceeds in ``diluted'' free fall,
\begin{equation}
-v=[2m_c(1-\lambda_o^{-2})/x]^{1/2}\ ,\
a=[m_c/2(1-\lambda_o^{-2})x]^{1/2}\ ,\ \dot{m}=m_c\ .
\end{equation}
The magnetic field advected to the center assumes a
split-monopole topology,
\begin{equation}
b_z=\frac{a}{\lambda_o}\propto x^{-1/2}\ ,\ 
b_{r,s}=\frac{m_c}{\lambda_ox^2}\propto x^{-2}\ .
\label{eta0}
\end{equation}
These results, obtained using the monopole approximation
(eq. [\ref{mpsi}]), change little when one corrects for the deviations
from the exact expressions for $g$ and $b_{\r,s}$.  They differ from
the ideal-MHD results presented in LS in that {\em the collapse occurs
without discontinuities in the flow parameters or their
derivatives}. The robustness of this conclusion is manifested by the
fact that the solution curve in the $(x,-v)$ plane never approaches too
closely to the singular line.  As we noted in \S 1, the initial state
adopted by LS is characterized by $v_o=0$: as a result, their
integration from $x=+\infty$ toward $x=0$ encounters the singular
line. In order to form a point mass at the origin, they then need to
connect to one of the ``minus'' solutions defined in Shu~(1977).  When
they perform the integration in the monopole approximation, the
connection takes place with continuous flow parameters (but
discontinuous first derivatives) at the point where the singular line
crosses the axis $-v=0$. However, when they use more accurate
expressions for $g_\r$ and $B_{\r,s}$, the curve intersects the
singular line a little bit below the axis $-v=0$, where (according to
Fig.~2 in Shu~1977) no ``minus'' solutions exist. In this case the
connection to a ``minus'' solution can be achieved only through a
discontinuous jump across the singular line, and this is the weak
shock that LS invoke. Our solution demonstrates that such a shock is
not required when $v_o\neq 0$. This result is confirmed by the
flux-frozen collapse model presented in \S 3.3 of CK.
(As discussed in \S~1, numerical simulations of nonmagnetic collapse
have shown that the Larson-Penston 1969 solution, which also has
$v_o \neq 0$, provides an accurate approximation to the physical state
at the instant of PMF in unmagnetized clouds.)

\subsection{The Ambipolar-Diffusion Shock}

We now proceed to derive the solution for the realistic, nonideal case
(i.e., $\eta\neq 0$ for $x\geq 0$).  We continue to employ the
monopole approximation for the gravity and magnetic field, which we
subsequently justify both by verifying that the correction terms
remain small and by comparing the results with the CK numerical
simulations. One immediately realizes that the split-monopole field
topology (eq.~[\ref{eta0}]) that characterizes the ideal-MHD solution
near the origin cannot apply in the presence of ambipolar
diffusion. This is because, for the split monopole, $b_{\r,s}\propto
x^{-2}\gg b_z\propto x^{-1/2}$ as $x
\rightarrow 0^+$, so the diffusive term in the induction equation
(eq.~[\ref{induction}]) greatly exceeds the advective
term near the origin.  In practice, the outward diffusion
prevents the formation of the split monopole field configuration
in the first place, and a {\em different} solution is established
near the origin. This solution is characterized by the magnetic
flux being left behind while matter continues to fall into the
center and forms there a point mass that grows with time
according to equation (\ref{pointmass}). The
flow consequently turns into an ``undiluted'' free fall at small $x$,
\begin{equation}
-v=2a=(2m_c/x)^{1/2}\ ,\ \dot{m}=m_c\ ,
\label{va}
\end{equation}
with coresponding field components
\begin{equation}
b_z=b_{r,s} = \frac{2^{1/4}m_c^{3/4}}{\eta^{1/2}x}\ .
\label{bb}
\end{equation}
In this case, although the magnetic field grows with decreasing
distance from the origin, the magnetic force remains
negligible in comparison with the gravitational pull of the
central point mass. Equations~(\ref{va}) and (\ref{bb}) also
demonstrate the interesting fact that the local (differential) mass-to-flux
ratio $\lambda= a/b_z \propto x^{1/2}\rightarrow 0$ as
$x\rightarrow 0^+$. However, the ratio of the spatially integrated mass
and flux diverges as the origin is approached, consistent with
the physical picture of mass accumulating at the center after
decoupling from the magnetic field. 

Under the flux-freezing conditions of ideal MHD, $\lambda(x) =  {\rm
const.}$ (see \S 3.3). It may at first seem
counterintuitive that, in the nonideal case,
the local mass-to-flux ratio decreases with decreasing $x$,
since one could expect ambipolar diffusion to become
progressively more efficient at loading magnetic field lines
with matter as one approached the densest parts of the core near
the center. Indeed, this is what happens {\em before} point-mass
formation (e.g., CM94; CM95).
However, {\em after} PMF the center acts as a sink of
matter, so the magnetic field lines become increasingly depleted
of mass as one gets closer to the origin (although the
integrated mass-to-flux ratio $m/\psi$ does increase with
decreasing $x$).

Our numerical procedure for obtaining the solution in this case has
been to integrate the differential equations inward and outward from
an intermediate point near the origin, connecting them to the
free-fall solution (eqs. [\ref{va}] and [\ref{bb}]) at small $x$ and
to the pivotal state (eq. [\ref{initial}]) at large
$x$. Figures~2$e$--2$h$ show the derived solution for the parameter
set $\lambda_o=2.9$, $-v_o=1.0 c$, $\xi_{-17}=5$, and $\dot{m}_o=3$. These
parameter values are typical of the outer layers of a supercritical
core in the CK simulations
and are consistent with physical quantities deduced from observations of
protostellar cores (e.g., Crutcher et al. 1993, 1994).
It is seen that, while the solution is everywhere continuous, it
exhibits an abrupt change in the flow parameters at a certain value of
$x$. This transition, which is characterized by a continuous evolution of the
physical parameters, a neutral flow velocity that remains supersonic
throughout, and a faster deceleration of the ions than of the
neutrals, represents a {\em C-shock}
(e.g., Draine
\& McKee~1993, Smith \& Mac Low~1997).  A similar ambipolar
diffusion-induced C-shock has been found in the numerical simulations
of CK (see \S 4 for a detailed comparison between the self-similar
solution and the CK results). We identify the position of the shock
with the location of the abrupt drop in the ion velocity and determine
the shock speed from the motion of this transition relative to the
origin. The numerical integration yielded $x_{\rm sh}=v_{\rm sh}=0.3$
and $\dot{m}_c=5.9$, or, equivalently,
\begin{eqnarray}
r_{\rm sh} & = & 119\ \left(\frac{T}{10\ {\rm
K}}\right)^{1/2}\ \left(\frac{t}{10^4\ {\rm yr}}\right)\  {\rm
AU}\ ,\nonumber \\ 
u_{\rm sh} & = & 0.06 \ \left(\frac{T}{10\ {\rm
K}}\right)^{1/2}\ {\rm km}\ {\rm s}^{-1}\ ,\nonumber \\
\dot{M}_c & = & 9.3\ \left(\frac{T}{10\ {\rm
K}}\right)^{3/2}\ {\rm M}_{\odot}\ {\rm per}\ 10^6\ {\rm yr}\ .
\label{results}
\end{eqnarray}
As an aid in visualizing this self-similar solution, we present in
Figure~3 time sequences of the disk density and radial velocity during
the collapse up to the time when a 1.1~$M_{\odot}$ star is assembled
at the center.

It is worth emphasizing that the C-shock we have found is {\em required} by
ambipolar diffusion and {\em is not} the shock discussed in LS
(which, as we pointed out in \S 3.3,
is avoided altogether in an isopedic disk that has a nonzero
initial infall velocity). In the present (diffusive) case the
magnetic field cannot remain attached to the matter as mass
starts to accumulate in the origin, so the field decouples from
the gas as soon as a point mass appears at the
center.\footnote{This behavior can also be understood by
comparing the ambipolar diffusion and gravitational contraction
timescales, whose ratio after PMF decreases with diminishing distance from
the origin --- signaling the onset of dynamically important
ambipolar diffusion on small scales --- on account of the
decrease in the free-fall time that is brought about by the mass
accumulation at the center (see \S 1 and \S 3.3 in CK).}
The region of decoupled flux grows outward at a fraction of the speed
of sound and a C-shock develops at its outer boundary.  An
interesting and somewhat counterintuitive result is that the
accretion rate onto the central point mass is {\em smaller}
(albeit only slightly) in the
case with ambipolar diffusion than in the flux-frozen case (compare
Figs.~2$d$ and 2$h$). The reason is that, as we have noted,
the local mass-to-flux ratio decreases when the magnetic field decouples
from the matter. The corresponding increase in the magnetic
force slows down the inflow and thereby reduces the inflow
rate. Note that the two components of the
magnetic force, the field tension and the field pressure gradient, are
everywhere of comparable magnitude, so one cannot neglect
either of them in the analysis.

\section{Discussion}

It is of interest to compare our self-similar model results with the
detailed numerical simulations of CK. The model they present is of a
$\sim 5~M_{\odot}$ molecular cloud core with temperature $T=10~{\rm K}$.
Just prior to PMF ($t \rightarrow 0^-$) the accretion rate is $\dot{M}=5.4 \
M_{\odot}~\rm{per}~10^6~\rm{yr}$ and the mass-to-flux ratio at
the center is $\lambda=3.6$ --- note, however, that unlike the
self-similar model, neither $\dot{M}$ nor $\lambda$ are
uniform at that time in the model core of CK (see their Figs. 1$c$ and 1$h$).
Immediately after PMF the accretion rate rises to $9.4\
M_{\odot}~\rm{per}~10^6~\rm{yr}$, and it then decreases to $5.6 \
M_{\odot}~\rm{per}~10^6~\rm{yr}$ by the time a $1 \ M_{\odot}$
protostar forms at the origin (which occurs at $t = 1.5 \times 10^5~\rm{yr}$).
\footnote{This time is equal to $1.1~\tau_{gr}$, where $\tau_{gr}
= [r_{M_\odot}^3/G M_\odot]^{1/2}$ is the gravitational contraction
($\approx$ free-fall) time at the location $r_{M_\odot}$ within the
core, which contains $1~M_\odot$ at $t=0$. This result is similar
to that observed in the nonmagnetic collapse solutions of Shu (1977),
Hunter (1977), and Foster \& Chevalier (1993).}
Rapid ambipolar diffusion in the inner flux
tubes halts the inward advection of magnetic flux, which piles up and
propagates outward as a hydromagnetic disturbance. As the front of
piled-up flux moves out to larger radii, the magnetic field behind the
disturbance and the neutral--ion collisional coupling become strong
enough to affect the infalling neutrals, and a shock forms in the
core. The shock is of the C type (e.g., Mullan 1971; Draine 1980): the
infall speed of the neutrals (in the shock frame) remains supersonic
while the infall speed of the ions is much smaller than the ion Alfv\'{e}n
speed. By the time a $1\ M_{\odot}$ protostar has formed at the
origin in their typical model calculation, the
shock is located at a radius $\sim 3.5 \times 10^3~\rm{AU}$ and
propagates outward with a speed (in the stellar rest frame) $\sim
0.7~C = 0.13~{\rm{km}}~{\rm s}^{-1}$. (For comparison,
eq.~[\ref{results}] yields $M_c\approx 1.4\ M_{\odot}$, $r_{sh}\approx
1.8 \times 10^3~\rm{AU}$, and $u_{sh}\approx 0.06~{\rm{km}}~{\rm
s}^{-1}$, respectively, for the accumulated central mass, shock
location, and shock speed at that time.)  As the hydromagnetic disturbance
propagates outward, the postshock accretion rate decreases due to the
fact that the neutrals are ``hung up'' (i.e., decelerated) in the
region of amplified magnetic field behind the shock front (see
Fig. 6$g$ of CK). This is also what happens in the self-similar model
(Fig. 2$h$).  As discussed in CK (see also Li
\& McKee 1996), the postshock region is potentially susceptible to
interchange instabilities (i.e., the gravitationally-induced interchange
of magnetic flux tubes; e.g., Spruit \& Taam 1990; Spruit,
Stehle, \& Papaloizou 1995): this issue, however, cannot be fully
addressed until nonaxisymmetric collapse simulations are performed.

In the limit $r \rightarrow 0$ (or $x \rightarrow 0$) the column density
and the infall speed of the neutrals are both $\propto r^{1/2}$
($\propto x^{1/2}$, see eq. [\ref{va}] and Figs. 2$e$, 2$f$,
3$a$, and 3$b$), reflecting the ``undiluted'' free-fall
collapse induced by the central point mass.
A similar behavior is also revealed in the numerical simulations
(see Figs. 6$c$ and 6$e$ of CK).

As mentioned in \S 3.4, ambipolar diffusion ``unloads'' the mass
in the effectively stationary (see Fig. 2$e$) flux tubes behind the shock
front. As a result, the local mass-to-flux ratio
$\lambda(x) = a(x)/b_z(x)$ is no longer a monotonically
decreasing function behind the shock (see Figs. 2$f$, 2$g$, and 3$b$);
this behavior is also seen in the CK simulations (see their Fig. 8$a$).

One obvious advantage of the simple self-similar model is that one can
scan the space of model parameters much more easily than is possible
in the case of the time-consuming numerical simulations. For
instance, the typical model presented in CK took $\sim$ two weeks of
computer time (running in background on an SGI R-4000 Indigo
workstation) for the central mass to accumulate $1 M_\odot$ of
material. By contrast, calculation of our self-similar models is
usually completed within half a minute (for typical models) when run
on the same workstation.  

We note that Li \& McKee (1996) suggested that Ohmic dissipation [a
process that,
for the assumed cloud composition and cosmic ray ionization rate,
becomes important for densities $n \ga
10^{11}~\rm{cm}^{-3}$ (e.g., Nakano \& Umebayashi~1986a,b),
or, equivalently, on 
scales $r \ll 1~\rm{AU}$]
would halt the accretion of flux onto a protostar, with the flux
consequently presenting
an obstacle to the neutrals and thereby giving rise to a shock.  However, as
shown by CK and verified by our self-similar model, ambipolar diffusion is
sufficiently efficient at much lower densities (or much larger radii) to
halt flux advection and cause the formation of a hydromagnetic shock,
independent of the effect of Ohmic dissipation.  Despite the
misidentification of the shock formation mechanism and the fact that
the induction equation was not solved for the magnetic
field structure, the simplified model
of Li \& McKee was found by CK to provide a reasonably good
approximation to the results of their numerical simulations.

As we have already noted in \S 3.1, the numerical results of
core collapse and point-mass formation presented in CK can be
described by a self-similar model only in an approximate sense. In
particular, the condition of the spatial uniformity of $u$,
$\dot{M}$, and $\lambda$ at the instant of PMF,
as well as the scaling $n_i \propto n^{1/2}$, are
generally not strictly
valid in the cores of the models presented in CM94, CM95,
BM94--BM95b, and CK.
Furthermore, the temporal behavior of the
physical quantities in our self-similar solution deviates from
that in CK. For example,
CK find that in the region of free-fall collapse
near the central point mass $-0.76 < d \ln \sigma/ d \ln t < -0.55$,
which is different from the self-similar result $d \ln \sigma / d \ln
t = -1/2$ (see eqs.~[\ref{nondimensional}] and [\ref{va}]). It
is worth stressing, however, that, in spite of these differences,
the self-similar solution accurately
describes the formation of an outward-propagating
ambipolar-diffusion shock as well as the other basic characteristics of the
evolution of a protostellar core during the post-PMF epoch.
As discussed in CK, the numerical simulations of
this model are consistent with observations of accreting protostars.
In particular, the inferred evolution of the mass accretion
rate is similar to that deduced for the protostellar objects HL Tauri
and L1551-IRS5, and the calculated magnetic field structure agrees
with polarimetric observations of dense protostellar cores (see \S~4.2
of CK).

\section{Conclusion}

In this paper we have presented a self-similar solution of the
collapse of a magnetized molecular cloud core (assumed to also
be nonrotating and isothermal) that, for the first time,
incorporated the effects of ambipolar diffusion in a
self-consistent manner. We have focused on the post-PMF
(point-mass formation) phase of the collapse of a disk-like
core, noting that Basu
(1997) had previously explored the self-similar nature of the
collapse before a central mass (i.e., a protostar) first appears
at the origin. We clarified the distinction between the ideal
and nonideal MHD cases by plotting the singular lines in the
position--velocity space and showing that
they correspond to different critical speeds (the magnetosonic
speed and thermal sound speed in the ideal and nonideal problems,
respectively). We obtained a solution for the ideal
(flux-frozen) case that exhibits a split-monopole field
topology near the center. This solution differs from
the one obtained by Li \& Shu (1997) in that it involves no
shocks. We showed that the shock in the LS solution is a direct
consequence of their assumption that the core at the time of PMF
is described by a stationary density distribution (corresponding
to a singular isothermal toroid), and we pointed out that a shock
will generally {\em not} be present under the more realistic
assumption of a nonzero inflow speed near the origin at that instant.
We demonstrated, however, that a shock is a generic feature of
the solution in the nonideal (ambipolar diffusion) case. This
(C-type) shock is a direct consequence of the action of ambipolar diffusion
in the central region of the core following PMF: the magnetic
diffusivity decouples the field from the matter, causing the
gas to free-fall to the center (where it accumulates in a point
mass) and the field to stay behind and drive a shock outward. 
We have compared this solution with the results of the numerical
simulations of Ciolek \& K\"onigl (1998) and confirmed that,
while the more realistic numerical models are not strictly self-similar,
our simplified solution nevertheless captures the main
features of the core evolution after PMF.

\acknowledgements{This work was supported in part by NASA grants
NAG 5-2766 and NAG 5-3687. Helpful comments by the referee, Prudence
Foster, are gratefully acknowledged.}

\newpage
\section*{Figure Captions}

{\bf Fig. 1.}---A plot of the function ${\cal R}(X)$, which
determines how different disk radii are weighted in the expressions for
the gravitational acceleration (eq.~[\ref{gravity}]) and
the surface radial magnetic field (eq.~[\ref{Brb}]). The
function is normalized by $\int_0^{\infty}{\cal
R}(X){\rm d}X=1$.

{\bf Fig. 2.}---Self-similar collapse solutions for an ideal-MHD
($\eta = 0$) disk ($a)-(d$) and for an ambipolar
diffusion-dominated ($\eta = 0.3$) disk ($e)-(h$), for initial
conditions represented by the parameters $v_o=-1.0$, $\lambda_o=2.9$, and
$\dot{m}_o=3$. Plotted as a function of the similarity variable
$x$ are the radial speeds (in units of the speed of sound $C$) of the neutrals
($v$, {\em solid curve}) and ions ($v_i$, {\em dashed
curve}) in panels ($a$) and
($e$), the dimensionless surface density $a$ in panels ($b$) and
($f$), the normalized vertical ($b_z$, {\em solid curve}) and
radial surface ($b_{r,s}$, {\em dashed curve}) magnetic field
components in panels ($c$) and ($g$), and the mass accretion rate
into the center ($\dot{M}$, for a temperature $T=10\ {\rm K}$)
in panels ($d$) and ($h$). Note that the ions comove with the
neutrals in the ideal-MHD case (panel [$a$]). The singular curves
({\em open circles}) in panels ($a$) and ($e$) correspond to
straight lines in nonlogarithmic units
[$x-v=(1+2\lambda_o^{-2})^{-2}$ in the ideal-MHD case and $x-v=1$
in the presence of ambipolar diffusion].  Both solutions are
continuous and have supersonic neutral speeds for all $x$. The abrupt ion
deceleration in the nonideal solution gives rise to a strong
C-shock. No shock appears in the ideal solution.  At large $x$
(equivalently, large $r$), $v\rightarrow {\rm const.}$ and $a\propto
b_z\approx b_{\r,s}\propto x^{-1}$.  At small $x$ (equivalently,
small $r$), the diffusive solution differs from the ideal-MHD one.
Both solutions contain a point mass at the center and therefore both
exhibit free-fall--type profiles ($v\propto a\propto x^{-1/2}$) near the
origin. However, the nonideal solution has $b_z\approx b_{r,s}\propto
x^{-1}$, whereas the ideal-MHD one ($b_z\propto a$) contains a split
magnetic monopole at the center ($b_z \propto x^{-1/2}$ and
$b_{r,s}\propto x^{-2}$).

{\bf Fig. 3.}--- The distributions of the core infall
speed $u$ and surface number density $N$ at times $t=0$, 2, 4, 6,
8, 10, and $12\times 10^4\ {\rm yr}$ {\em after} a point mass forms
at the center. By the end of the displayed evolution, 1.1 solar
masses have accumulated at the origin.

\begin{figure}
\plotone{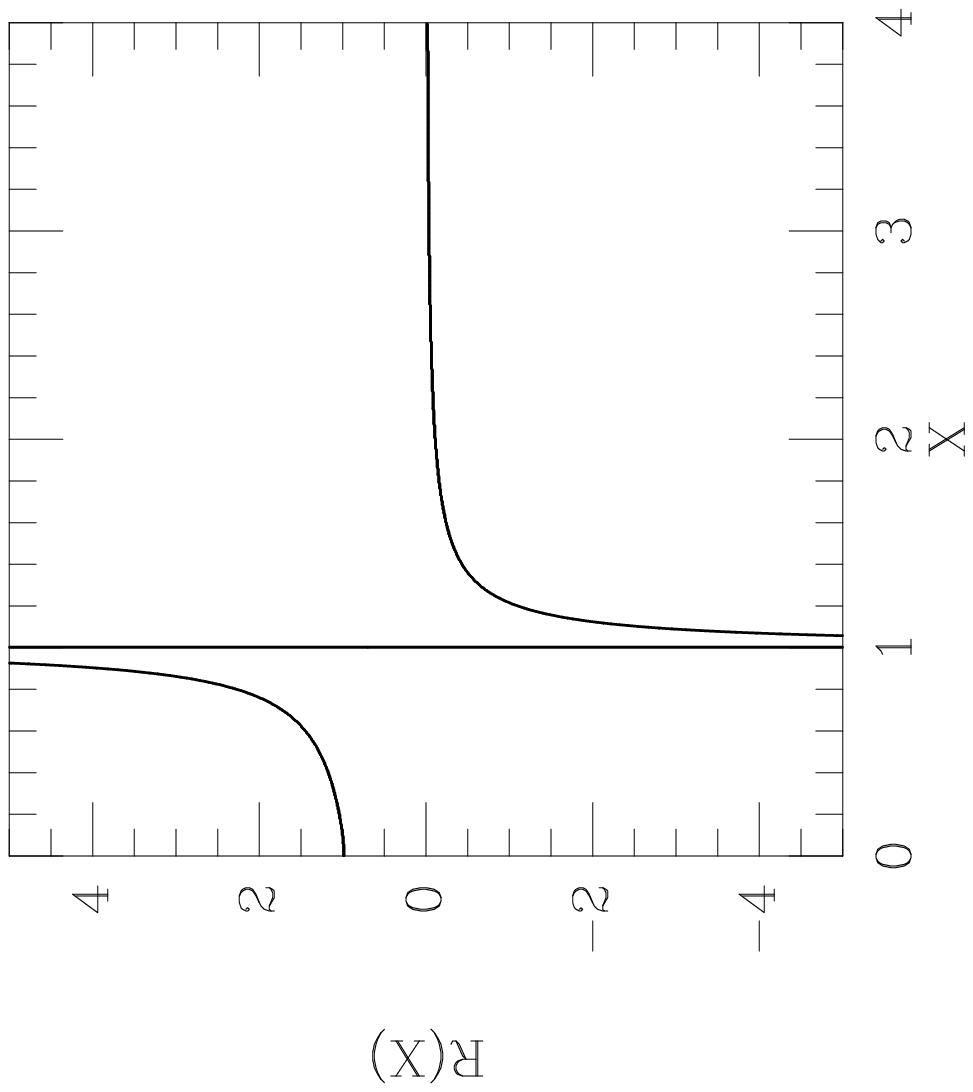}
\caption{}
\end{figure}

\begin{figure}
\plotone{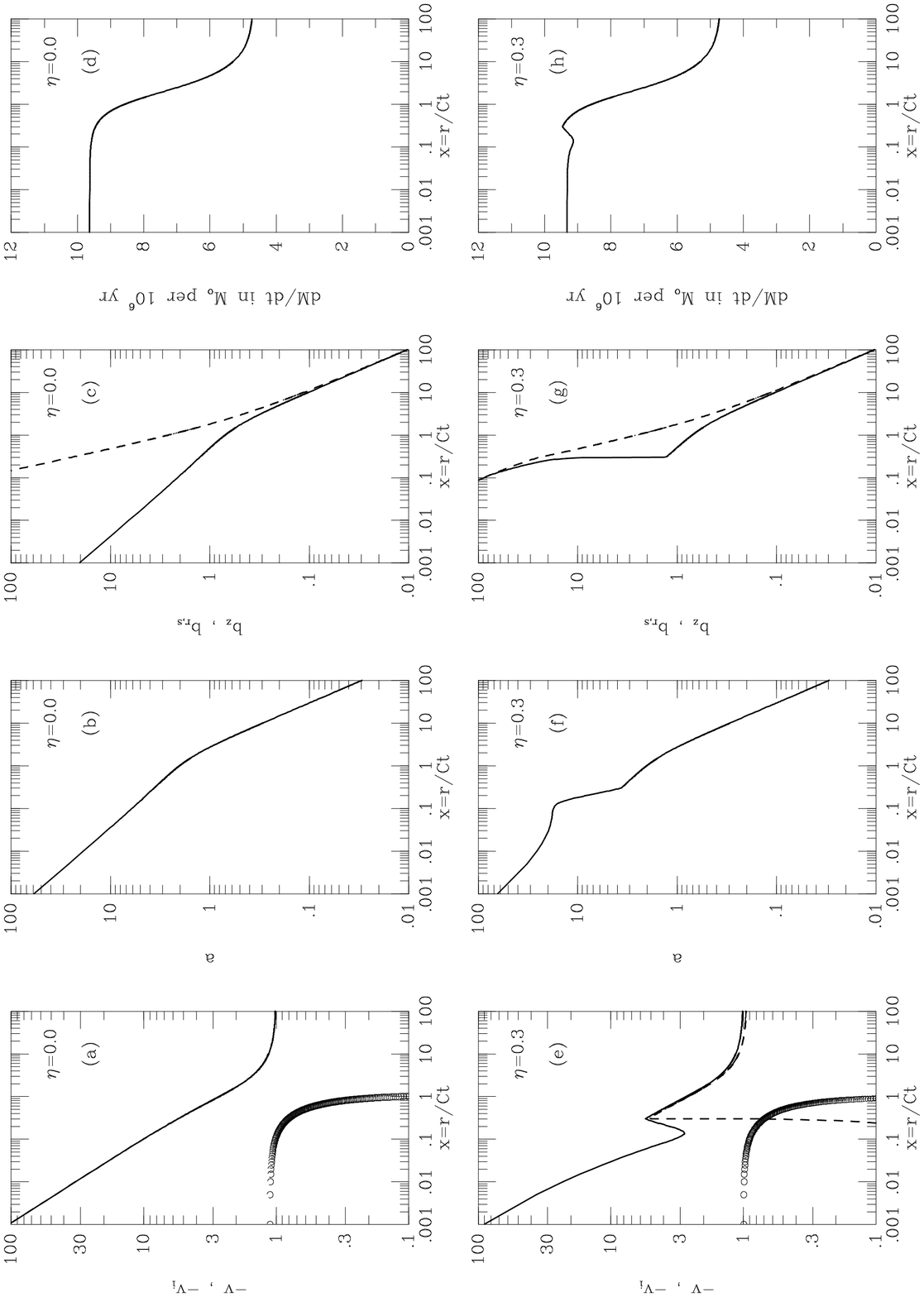}
\vspace{3ex}
\caption{}
\end{figure}

\begin{figure}
\plotone{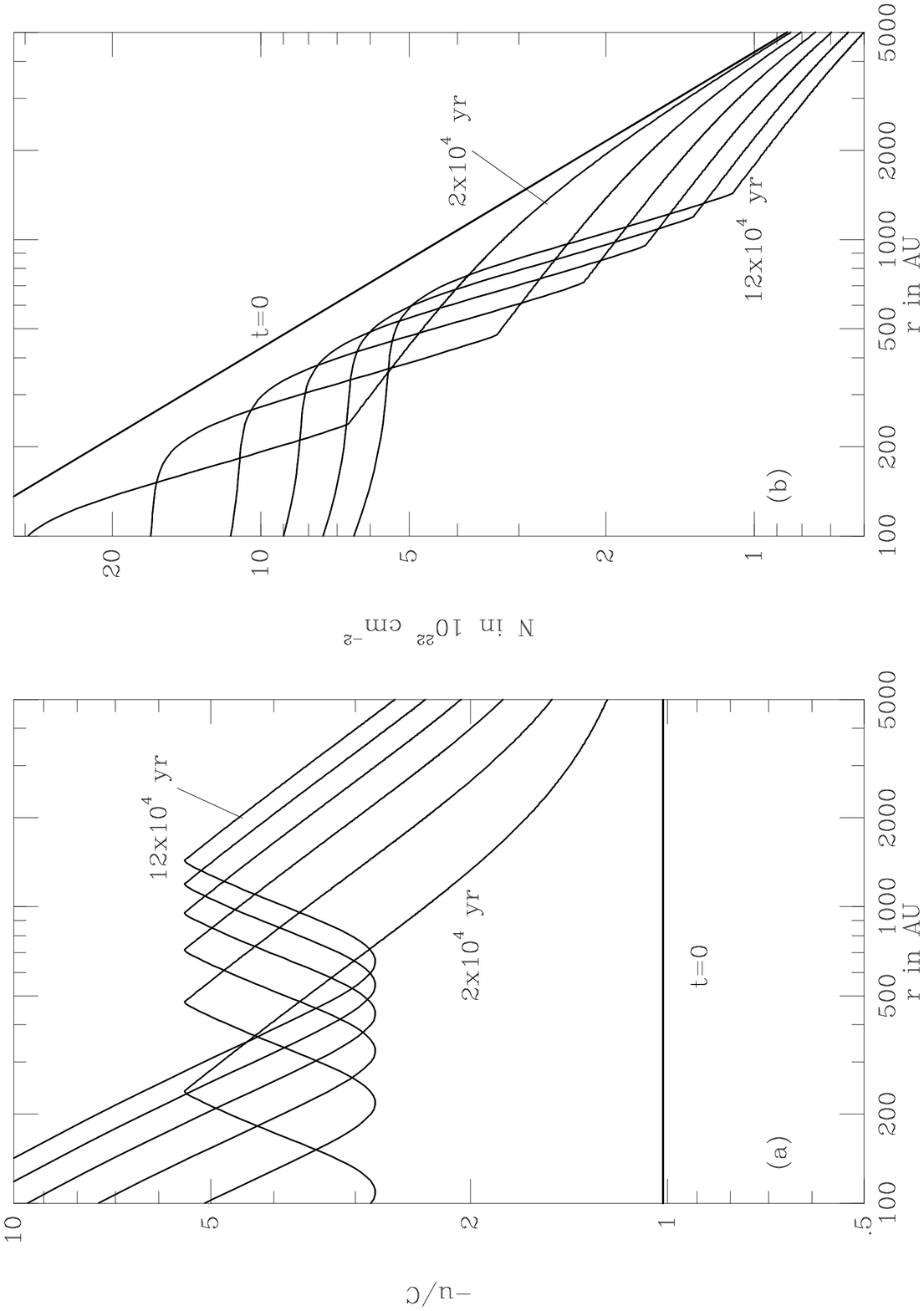}
\caption{}
\end{figure}

\end{document}